\documentstyle[12pt]{article}

\newcommand{\postscript}[2]{\vspace{5cm}}

\begin{document}
\newcommand{\beq}{\begin{equation}}
\newcommand{\eeq}{\end{equation}}
\newcommand{\beqa}{\begin{eqnarray}}
\newcommand{\eeqa}{\end{eqnarray}}
\def\lag{Lagrangian}
\def\ra{\rightarrow}
\def\RA{\rightarrow}
\def\x{\times}
\def\etal{{\it et al.,\ }}
\def\sinn{\sin^2 \theta_W}
\newcommand{\RMP}[3]{{Rev. Mod. Phys.} {\bf #1}, #2 (19#3)}
\newcommand{\PR}[3]{{Phys. Rev.} {\bf #1}, #2 (19#3)}
\newcommand{\PL}[3]{{Phys. Lett.} {\bf #1}, #2 (19#3)}
\newcommand{\Rep}[3]{{Phys. Rep.} {\bf #1}, #2 (19#3)}
\newcommand{\Ann}[3]{{Ann. Rev. Nucl. Sci.} {\bf #1}, #2
(19#3)}
\newcommand{\NS}[3]{{Nucl. Sci.} {\bf #1}, #2 (19#3)}
\newcommand{\PRL}[3]{{Phys. Rev. Lett.} {\bf #1}, #2 (19#3)}
\newcommand{\NP}[3]{{Nucl. Phys.} {\bf #1}, #2 (19#3)}
\newcommand{\con}[3]{{\bf #1}, #2 (19#3)}
\def\mev{\; {\rm MeV} }
\def\MEV{\; {\rm MeV} }
\def\Ev{\; {\rm eV} }
\def\tev{\; {\rm TeV} }
\def\eV{\; {\rm eV} }
\def\gev{\; {\rm GeV} }
\def\etc{{\it etc}}

\begin{center}

{\bf FIVE PHASES OF WEAK NEUTRAL CURRENT EXPERIMENTS FROM THE
PERSPECTIVE OF A THEORIST\footnote{Invited talk presented at
{\it 30 Years of Neutral Currents. From Weak Neutral Currents to the
(W)/Z and Beyond}, Santa Monica, February 1993.}}

\vspace{3ex}

Paul Langacker \\ University of Pennsylvania \\ Department of Physics
\\ Philadelphia, Pennsylvania, USA 19104-6396
\\  \today, UPR-0553T,  hep-ph/9305255

\vspace{6ex}

Abstract

\end{center}

I give my personal perspective on the past, present, and future
of weak neutral current experiments, emphasizing the experimental
inputs; the theoretical difficulties and inputs; the role of
model independent and global analyses; and the implications at
each stage.

\vspace{6ex}

\begin{itemize}
\item Introductory Comments
\item The Discovery Phase (unification)
\item The Second Generation (the standard model confirmed)
\item The Third Generation  (precision tests; radiative corrections)
\item The LEP Era (high precision; $m_t$; new physics searches)
\item The Future (complement to colliders)
\end{itemize}

\newpage

\section{Introductory Comments}

I have been interested in the implications of weak neutral currents
for some 17 years.  In this talk I will describe the five major phases
of experiments from my own theoretical point of view.  Let me begin
with some general comments.

\begin{itemize}

\item The weak neutral current (by which term I include the
properties of the $W$ and $Z$ bosons) has always been the primary
test of the electroweak unification. QED and the weak charged
current theory already existed and were incorporated into the
standard model.  (The latter was, however, greatly improved by
the possibility of computing radiative corrections.)

\item The weak neutral current experiments have uniquely
established the fermion couplings, consistent with the gauge
group and fermion representations of the standard model.

\item The weak neutral current has also probed the underlying
structure of gauge field theory.  The electroweak unification in
itself is a significant success of the gauge concept.  Furthermore,
the precision experiments require the calculation of radiative
corrections, which tests the whole concept of gauge invariance
and renormalization theory.  The electroweak unification has also
made possible the calculation of finite higher-order corrections
to weak charged current processes, which are essential for
agreement between theory and experiment.

\item Another role has been the search for new physics.  So far
all data are in agreement with the standard model.  Moreover,
there are large domains of possible new physics which are
excluded.  I expect that this role will continue to be
significant in the future, and that precision experiments will be
a useful complement to high energy colliders.

\item I would also like to emphasize the complementarity of the
various precision experiments.  No one experiment is sensitive to
every type of new physics or to every aspect of the standard
model.  However, because of the wide variety of reactions and
kinematic ranges that have been explored it is unlikely that any
relevant type of new physics would be able to slip through
without leaving a signature.

\item The program has been aided by the global
analysis of all experiments simultaneously.  A global analysis
has the advantage that the experiments collectively contain more
information than any one, but has the obvious caveat that one
must be careful in the estimation and application of experimental
and theoretical uncertainties and their correlations.

\end{itemize}

\section{The Five Phases}

Let me first give a brief overview of the five phases.

\begin{itemize}

\item The discovery phase, which culminated in 1973 with the
discovery of the weak neutral current, has been extensively
discussed at this conference.  In particular, the existence of
the weak neutral current was a successful prediction of the $SU_2
\x U_1$ model.

\item The second generation of experiments occurred in the latter
half of the 1970's, and was dominated by purely weak $\nu N$ and
$\nu e$ scattering, and by weak-electromagnetic interference in
the polarized $e^{\uparrow \downarrow}D$ asymmetry from SLAC.
These experiments were sufficiently precise (typically $\sim$10\%)
that it was possible to begin model-independent fits, which means
an analysis allowing for an arbitrary gauge theory.  It was
possible to determine most of the vector and axial parameters of
the four-Fermi interactions and to show that they were consistent
with the $SU_2 \x U_1$ model to first approximation, and not some
completely different theory.  Assuming the standard model, one
was able to obtain a reasonably precise value for the weak angle,
$\sin^2 \theta_W = 0.229 \pm 0.010$, where the error was mainly
experimental.  Although it was not presented in this way, the
results implied an upper limit of $m_t < 290 \gev$.

\item The third generation of experiments, in the 1980's, was
characterized by higher precision (typically 1 -- 5\%) and the
existence of many more probes.  These included purely weak $\nu
N$ and $\nu e$ scattering, as well as a number of
weak-electromagnetic interference phenomena.  New results
included atomic parity violation, $e^+e^-$ annihilation, and the
actual observation of the $W$ and $Z$ bosons at CERN with a
determination of their masses.  The result was that the standard
model was correct, and complicated alternative models (with
similar four-Fermi interactions) were excluded.  Furthermore, for
the first time the weak interactions of the $b$ quark were
measured in both charged and neutral current processes, with the
consequence that the $t$ exists, and searches were made for and
limits set on many types of new physics.  A more precise value of
the weak angle was obtained, $\sin^2 \theta_W = 0.230 \pm 0.007$,
where now the error was mainly theoretical from the
interpretation of deep inelastic scattering.  One obtained a more
stringent limit $m_t < 200 \gev$.

\item The fourth phase is the LEP era, which began in 1989.  This
is dominated by the $Z$-pole observables including $M_Z$ and the
$Z$ widths and asymmetries, which are typically at the few tenths
percent level.  There have also been a number of other
experiments, including much improved measurements of $M_W$,
atomic parity violation (APV), and more precise $\nu e$
scattering.  They test the standard model at the level of
radiative corrections and stringently search for new physics.
The weak angle is now determined an order of magnitude better
than before, $\sin^2 \hat{\theta}_W (M_W) = 0.2328 \pm 0.0007$,
where now the error is almost entirely due to the uncertainty in
$m_t$.  One also has the standard model prediction $m_t =
150^{+19+15}_{-24-20} \gev$, where the second error is from the
Higgs mass.

\item Finally, there is the possibility of future ultrahigh
precision $(\ll 1$\%) experiments.  These include polarization
asymmetries, much improved atomic parity violation experiments,
determinations of $M_W$, and a possible $\nu N$ experiment at
Fermilab.  These experiments would be sensitive to many types of
new physics up to the TeV, range and would be a useful complement
to the direct searches for new particles at the SSC and LHC.

\end{itemize}

\section{The Discovery Phase}

I don't have much to say about this phase from personal
experience.  (I was working on completely different things at the
time.)  Nevertheless, I would like to make a few comments.

\begin{itemize}

\item It is important to recall the historical context.
The discovery of the weak neutral current occurred
during the period of -- and was in part responsible for -- a
complete change of outlook in particle physics.  Previously, the
emphasis had been on the classification of elementary particles
and their properties, and most effort was devoted to the strong
interactions and $S$-matrix theory.  Gradually, however, the
framework changed to quantum field theory.  This change was
effected by theoretical developments, the parallel development of
quarks and QCD, and of course the discovery of the weak neutral
current.

The original role of the flavor-changing neutral currents (FCNC)
can be characterized as looking under the wrong lamppost.  As we
have heard at this meeting, the severe limits on the FCNC
confused the issue because it was reasonable to expect that if
they were small then the flavor-conserving interaction should
also be small.  Furthermore, their absence delayed the
theoretical development of the hadronic part of the standard
model, which required the GIM mechanism~\cite{GIM}.  But this
needed the charm quark, which people were reluctant to accept
until the discovery of the $J/\psi$ and neutrino-induced dimuons.
FCNC are now very important again,
because they are predicted at some level by most extensions of the
standard model.  In particular, theories involving compositeness
and/or dynamical symmetry breaking generally predict FCNC far in
excess of the existing limits.  The experimental limits have
severely held up the development of realistic models, and perhaps
even cast doubt upon that whole set of ideas.  It is important to
push the searches as far possible, not only in the kaon system,
but also in heavy quark and lepton decays.

\item The weak neutral current has always been closely connected
with the unification of the weak and electromagnetic
interactions.  Almost all unified theories predict neutral
currents, and, conversely, the most natural way to have neutral
currents is within gauge unification.  Furthermore, neutral
currents (or the ratio of neutral current and charged current
rates) have always been the primary quantitative test of the
structure of the standard model and have been crucial in the
search for new physics.

The Fermi theory of the weak charged current and QED existed
before the unification.  They were incorporated, and the Fermi
theory was improved in the sense that unification made
possible the calculation of higher-order corrections.
Nevertheless, the weak charged current has so far been less
important in establishing the standard model and searching for
new physics.  This is due in part to the fact that charged
current experiments tend to have more hadronic uncertainties.
Furthermore, precise measurements often result in the measurement
of an element of the CKM mixing matrix \cite{CKM} rather than
directly testing the standard model.  Of course, verification
that the CKM matrix is unitary is important.  For example, the
successful relation \cite{univ}
\beq | V_{ud}|^2 + |V_{us}|^2 + |V_{ub}|^2 = 0.9992 \pm 0.0014 \eeq
eliminates many extensions of the standard model involving
right-handed currents or extended fermion sectors.  This success
is particularly remarkable in that if one did not apply radiative
corrections to $\beta$ and $\mu$ decay one would have obtained
$1.036$, in contradiction with unitarity.  Without the entire
apparatus of non-abelian gauge theory one would not have been
able to calculate these corrections, which would have been
infinite and meaningless, so (1) tests the underlying ideas of
gauge theory.  In the future we expect other important
probes of CKM unitarity, especially CP violation in $B$ decays.
Nevertheless, the neutral current has so far been the most
important test of the standard model.

\item As we have heard at this meeting there were many early
confusions.  Weak interaction experiments are hard, and incorrect
results are sometimes obtained.  There is a long history of this
in charged current interactions, neutrino mass, \etc., and the
neutral current is no exception.  Incorrect early results
significantly delayed the discovery of the weak neutral current.
Even after it was found the confused situation for the first few
years, such as unsuccessful searches for
atomic parity violation, led to a plethora of alternative
theoretical models.  This takes me to the second phase.

\end{itemize}
\newpage

\section{The Second Generation}

The second generation of experiments in the second half of the
1970's, which were typically of 10\% precision, clarified the
situation.  In particular, they established that the basic
structure of the standard model was correct, at least for the
4-Fermi interactions relevant at low energies, and many
alternative models with disjoint parameters were eliminated
\cite{KIM}.

\subsection{The Experiments}

Most of the experiments were purely weak processes involving the
neutral current scattering of neutrinos.  The most precise were
the CERN and Fermilab measurements of deep inelastic scattering,
$\nu N \ra \nu X$, from targets that were approximately
isoscalar.  Although these gave the most precise results, if one
wanted to independently determine the isospin structure of the
neutral current one needed other information, such as could be
provided by deep inelastic scattering from $p$ and $n$ targets,
elastic scattering $\nu p \ra \nu p$, and inclusive and exclusive
pion production $\nu N \ra \nu \pi X$, $\nu \pi N$.  There were
also several (low statistics) measurements of $\nu_\mu e \ra
\nu_\mu e$.  This era also featured the seminal SLAC polarized
$e^{\uparrow \downarrow} D \ra eX$ asymmetry
measurement~\cite{SLAC}, which established parity
violation in the weak neutral current.  The first successful
atomic parity violation experiments in bismuth and thallium were
reported towards the end of this period.  There had been several
early incorrect null experiments, as well as considerable
theoretical difficulty in the interpretation; therefore, atomic
parity did not play a significant roll in this phase.  (The null
experiments made even more significant the SLAC
asymmetry measurement.)

\subsection{Theoretical Inputs}

Due to the variety of competing models it was important to have a
general way of analyzing the data to distinguish between them.
Each model had its own set of parameters and it was awkward to
describe experimental results in terms of the parameters of each
of several models.  Therefore, the idea of model independent
analyses of 4-Fermion couplings was devised, and was especially
emphasized by Bjorken and by Hung and Sakurai~\cite{modind}.  One
utilizes a parametrization in which each 4-Fermi interaction
allows arbitrary admixtures of $V$and $A$ coefficients for the
electrons and quarks (one assumes a $V-A$ coupling for the
neutrinos), as is valid for an arbitrary gauge theory.  Each
individual gauge model would give specific predictions for these
coefficients.  One then attempts to determine them from
experiment in a model independent way to see which models are
allowed or excluded.

$S$, $P$, and $T$ couplings are not generally included in the
model independent analyses.  Even today, one could not rigorously
exclude significant amounts of $S$, $P$, and $T$ from the $\nu
N$ and $\nu_\mu e$ reactions.  However, they are rendered largely
superfluous by the successes of the standard model and the
discovery of the $Z$ boson.  Furthermore, the SLAC asymmetry and
(in later generations) other measurements sensitive to
weak-electromagnetic or neutral-charged current interference
excluded the possibility of a dominant role for $S$, $P$, and
$T$.  In the interest of simplicity and for lack of theoretical
motivation, we will therefore ignore the possibility of $S$, $P$,
and $T$ except as a small perturbation.

In the mid 1970's there were reported anomalous trimuon events by
the HPW $\nu N$ experiment at Fermilab.
Although these did not ultimately
survive, they stimulated several groups to develop alternative
gauge models to account for them.  I was guilty of some of this
work myself, and my interest in neutral currents came about from
looking for ways to constrain or test these alternative models.
By that time the experiments were getting quite accurate, and I
decided that it would be useful to gather all of the neutral
current data and carry out a model independent analysis.
For this one needs enough
constraints to determine all of the couplings, and this in turn
requires a simultaneous (global) analysis of all of the data.
Another thing that was needed at that stage was a better
theoretical treatment of the reactions.  In particular, the deep
inelastic data was becoming sufficiently precise that simple
parton model expressions were no longer adequate.  It became
clear to my collaborators and me \cite{KIM}
that it would be necessary to
apply QCD corrections, $c$ thresholds, \etc., uniformly to all of
these experiments to obtain reliable theoretical formulas from
which to extract information about the weak interactions.

\subsection{A Digression: Global Analyses of Data}

The application of global analysis techniques is now well
accepted, but was somewhat controversial at the time of the
second generation of experiments.  A global analysis is basically
the combination of two or more measured numbers, which may be
obtained in the same or different experiments, to obtain a
result.  A global analysis often contains more information than
any one experiment, but care must be taken with uncertainties.

Often one applies a global analysis without using that language.
For example, in QED one obtains $\alpha$ to a very high
precision from the quantum Hall effect.  However, that by itself
does not test QED.  To do so one must simultaneously analyze one
or more other measurements, such as of the muon anomalous magnetic
moment, $g_\mu -2$.  QED is tested by their consistency, {\it
e.g.}, by using the result of one experiment to predict another
within the QED framework.

In the standard model weak neutral current sector the results of
any one experiment can usually be accommodated by choosing the
value of the weak angle $\sin^2\theta_W$.  However, the large
(unknown) value of $m_t$ means that it plays a significant role
in the radiative corrections.  Therefore, any complete extraction
of standard model parameters requires at least two measurements.
However, even here one is not really testing the standard
model\footnote{There may, of course, be ranges of experimental
values that cannot be described by {\em any} values of
$\sin^2\theta_W$ and $m_t$.}: one needs to determine these two
parameters and then predict the results of a third experiment.
In fact, one wants to have as many different measurements as
possible.  Each experiment has different dependences on the
parameters and on the various types of new physics, and one wants
to maximize the sensitivity.  Of course, if any deviation is
observed one would want to have as many independent probes as
possible to confirm it and to diagnose its origin out of the
enormous variety of possible types of new physics.

Another advantage of a global analysis is that it allows one to
determine the parameters for larger classes of models, such
as arbitrary gauge theories.  Once the generalized parameters are
determined one can see whether the standard model is uniquely
selected and set limits on small perturbations around it.

Finally, there are often different experiments measuring similar
things, such as deep inelastic scattering.  It is important not
only that one use the best possible theoretical expressions in
the analysis but that they be applied uniformly.  For example, it
is desirable to use the same quark distribution functions for
each experiment, and crucial that common theoretical
uncertainties are properly correlated.

On the other hand, there are obvious dangers when one combines
experiments.  In particular, one must be careful with
systematic and theoretical uncertainties and the correlations
between measurements.
Because of the importance of careful error estimation and
combination I would like to emphasize what I
consider the four mortal sins of data presentation:

\begin{itemize}

\item The first is to underestimate systematic or theory errors.

\item The second, and almost as serious, is to overestimate
systematic or theoretical errors.  The old idea of multiplying an
uncertainty by $\pi$ may be reasonable for giving an absolute
bound on the range of errors, but it can be misleading if one
tries to use the quoted uncertainty quantitatively.  Based
on my own observation of experiments and how they have
compared with later more precise results, I suspect that
there has been a tendency to overestimate systematic errors.

\item A third mortal sin is to not publish error
correlation matrices.

\item Finally, it is important to not present experimental
results in too narrow or trendy a context.  For example, it is
more useful to publish observables (such as cross sections) which
can be interpreted or analyzed in the context of a general gauge
theory (or, hopefully, an even more general framework), rather
than just $\sin^2 \theta_W$.  Such results can be interpreted in
a wider context, can be more readily used for setting limits or
searching for new physics, and can be later updated if the
theoretical calculations of strong interactions effects,
radiative corrections, \etc., are improved.  Obviously, detector
dependent-artifacts, acceptances, \etc., have to be corrected
for.\footnote{Such corrections are usually model dependent.
However, these may be calculated assuming the standard model,
which we now know to be an excellent first approximation.
Similar statements apply to many radiative corrections.} One can
present results both in this form and in terms of
$\sin^2\theta_W$.

\end{itemize}

\subsection{Results}

There were several global analyses of the second generation of
the experiments.  I will report on those in the article by Kim
{\it et al.} \cite{KIM}.

\begin{itemize}

\item Model independent analyses of the 4-Fermi couplings were
carried out.  The couplings relevant to $\nu q$ and $e q$
interactions were uniquely determined for an arbitrary $V$ and
$A$ theory (assuming $V-A$ neutrino couplings and family
universality), while the $\nu e$ couplings were determined up to
a two-fold ambiguity.  The results are shown in Figures~1, 2, and
3 and in Table~1.  Many alternative models with disjoint
predictions for these couplings were eliminated.

\begin{figure}
\vspace{10cm}
\caption[]{Allowed regions for the couplings relevant to
neutrino quark interactions in 1981, from \protect\cite{KIM}.  Only the
cross hatched regions are allowed by all of the data.  The left
and right chiral couplings to the up and down quarks are defined
by the effective interaction
$ -L^{\nu N} = \frac{G_F}{\sqrt{2}} \bar{\nu} \gamma_\mu (1 -
\gamma_5) \nu J^{\mu H},$ where $
J^{\mu H} = \sum_i \left[ \epsilon_L (i) \bar{q}_i \gamma^\mu (1 -
\gamma_5) q_i + \epsilon_R (i) \bar{q}_i \gamma^\mu (1 + \gamma_5) q_i
\right] .$
In the standard model the couplings are predicted at tree-level to be
$ \epsilon_L (u) \simeq \frac{1}{2} - \frac{2}{3} \sin^2\theta_W$,
$\epsilon_R(u) \simeq  - \frac{2}{3} \sin^2\theta_W,$
$\epsilon_L (d) \simeq - \frac{1}{2} + \frac{1}{3} \sin^2\theta_W,$ and
$\epsilon_R (d) \simeq \frac{1}{3} \sin^2 \theta_W$. The angles
$\theta_{L,R}$ are measured with respect to the vertical
($\epsilon_{L,R} (d)$) axes.}
\label{fig1}
\end{figure}

\begin{figure}
\vspace{9cm}
\caption[]{Leptonic coupling constants allowed at 90\%
confidence level in 1981,
from \protect\cite{KIM}.  The vector and axial vector couplings are
defined by
$ -L^{\nu e} = $ \\
$\frac{G_F}{\sqrt{2}} \bar{\nu} \gamma_\mu (1 -
\gamma_5) \nu \bar{e} \gamma^\mu \left( g^e_V - g^e_A \gamma_5 \right)
e.$
In the standard model,
$ g^e_V \simeq -\frac{1}{2} + 2 \sin^2\theta_W$,
$g^e_A \simeq - \frac{1}{2}$.}
\label{fig2}
\end{figure}

\begin{figure}
\vspace{15cm}
\caption[]{Regions allowed by the SLAC and early atomic parity
violation experiments in 1981,
from \protect\cite{KIM}.  The parity violating
couplings are defined by
$ L^{eq}_{PV} = \frac{G_F}{\sqrt{2}} \sum_i \left[ C_{1i} \bar{e}
\gamma^\mu \gamma_5 e \bar{q}_i \gamma_\mu q_i + C_{2i} \bar{e}
\gamma^\mu  e \bar{q}_i \gamma_\mu \gamma_5 q_i \right].$
    In the standard model,
$ C_{1u} \simeq -\frac{1}{2} + \frac{4}{3} \sin^2 \theta_W$,
$C_{1d} \simeq \frac{1}{2} - \frac{2}{3} \sin^2 \theta_W,$
$C_{2u} \simeq -C_{2d} \simeq-\frac{1}{2}+ 2 \sin^2\theta_W.$}
\label{fig3}
\end{figure}


\begin{table}
\begin{tabular}{|ccccc|}            \hline
  & 1980 & 1987 & present & SM ($M_Z$) \\ \hline
  $g_L^2$ & $0.296 \pm 0.008 $ & $ 0.2996 \pm 0.0044$ &
$ 0.3003 \pm 0.0039$ & 0.3021 \\
  $g_R^2$ & $0.032 \pm 0.007 $ & $ 0.0298 \pm 0.0038$ &
$ 0.0323 \pm 0.0033$ & 0.0302 \\
  $g_V^e$ & $ +0.043 \pm 0.063$
  & $-0.044\pm 0.036 $ & $ -0.035 \pm 0.017$ & $-$0.037 \\
  $g_A^e$ & $ -0.545 \pm 0.056$
  & $-0.498\pm 0.027 $ & $ -0.508 \pm 0.015$ & $-$0.506 \\
$ \sinn$& $0.229 \pm 0.010 $ & $0.230 \pm 0.007 $ & $ 0.2328 \pm 0.0007$
      & -- \\
       & (bare) &  (on-shell) & $\overline{\rm MS}$ &  \\
      $m_t$ & $< 290$ & $< 200$ & $ 150^{+19 +15}_{-24 - 20}$ & 150
(input)
       \\  \hline
\end{tabular}
\caption[]{Values of the $\nu q$ and $\nu e$ model
independent parameters, $\sin^2\theta_W$, and the $m_t$
predictions at various phases, compared to the standard model
prediction using as input $M_Z = 91.187 \gev$ and $m_t = 150
\gev$.  $g^2_{L,R}$ are defined by
$g^2_{L,R} = \epsilon_{L,R} (u)^2 + \epsilon_{L,R} (d)^2$.}
\label{tab1}
\end{table}

\item Assuming the standard model, one had a rather precise
value for the weak angle, namely $\sinn =0.229 \pm 0.010$.  Also
one had a result on the parameter $\rho_0 \equiv M_W^2 / M_Z^2
\cos^2 \theta_W$, which could be interpreted as an upper limit
$m_t < 290 \gev$.  The $\sin^2 \theta_W$ error was dominated by
experimental uncertainties, although the theoretical
uncertainties were not negligible.

At about the same time grand unified theories, {\it e.g.}, based
on the $SU_5$ model, became popular, which predicted $\sinn =
0.209^{+0.003}_{-0.002}$.  The experiments were
already precise enough to be problematic for these
models~\cite{GUT}, even before the nonobservation of proton
decay.  Within a year or two several groups~\cite{SUSY} pointed
out that supersymmetric extensions of the standard model, such as
${\rm SUSY}-SU_5$, led to a larger prediction
$0.225^{+0.015}_{-0.002}$, in good agreement with experiment.
This is illustrated in Figure~\ref{fig4}.

\begin{figure}
\vspace{9cm}
\caption[]{Predictions of the ordinary (open circles)
and supersymmetric (triangles and boxes) grand unified
models for $\sinn$, compared to the experimental data (filled
circles).}
\label{fig4}
\end{figure}

\item  There were also early limits on additional heavy $Z'$ bosons.
\end{itemize}

\section{The Third Generation}

The third generation of experiments in the 1980's were of the 1
-- 5\% level.  They not only made more precise the statement that
the standard model is correct to first approximation, but they
also ruled out many epicycle models which reproduced the low
energy 4-Fermi interactions but involved different $W$ and $Z$
masses.  These experiments were also the first to see the effects
of radiative corrections in a significant way.

\subsection{The Experiments}

This generation included high precision $\nu N$ and $\nu_\mu e$
experiments at CERN, Fermilab, and elsewhere, and also the first
$(\bar{\nu}_e) \nu_e e$ experiments at the
Savannah River reactor and at LANL.
The latter were sensitive to interference between the neutral and
charged currents.  Furthermore, there are many measurements of
weak-electromagnetic interference, from PEP, PETRA, and TRISTAN
below the $Z$-pole, including $e^+e^- \ra e^+e^-$, $\mu^+ \mu^-$,
$\tau^+\tau^-$, and hadrons.  A new generation of precise atomic
parity violation experiments in the cesium atom were performed in
Paris and Boulder.  Not only were the experiments much better,
but cesium is a simple atom with a single valence electron
outside a tightly bound core, allowing a clean calculation of the
relevant atomic theory.  There was also a $\mu C$ asymmetry
experiment at CERN.  Finally during this era the $W$ and $Z$ were
directly produced and observed at CERN and later at Fermilab, and
their masses determined.

\subsection{Theoretical Inputs}

The new higher precision required a careful attention to
radiative corrections and to the definitions of the renormalized
weak angle $\sinn$.  Marciano and Sirlin~\cite{MS} and others
carried out careful calculations of the corrections to all
relevant 4-Fermi processes.  Another necessary input was a more
careful evaluation of the theoretical formulas for deep inelastic
scattering.  This was greatly helped by an analysis of
Llewellyn-Smith~\cite{LLS}, who used isospin arguments to show
that most of the structure function uncertainties (other than
those associated with the $c$-quark threshold and non-isoscalar
targets) cancelled in the ratio of neutral and charged current
cross sections.  During the mid-1980's, I spent a great deal of
time reanalyzing all of the deep inelastic experiments to
estimate the remaining corrections and their uncertainties.  Each
experimental collaboration had done their own analysis, but had
usually only extracted $\sinn$.  For a model independent analysis
one needs a more general expression.  It was necessary to
reanalyze all of the experiments, folding in the appropriate
cuts, acceptances, and spectra.\footnote{It was just possible for
an outsider to reanalyze this generation of experiments.  That
could never be done with the more complicated LEP experiments.}\
The model independent reanalysis also allowed one to apply
uniform theoretical expressions to all of the experiments and to
properly correlate the theoretical uncertainties~\cite{amaldi}.

Another theoretical input was a parameterization of the effects
of certain types of new physics, such as $Z'$
bosons~\cite{durkin}, exotic fermions which mix with the ordinary
fermions~\cite{london}, \etc.  One needs an explicit
parametrization of how they affect each observable.  This should
be fairly general and not tied to a specific model, but on the
other hand there must be a small enough number of parameters to
be manageable.  One would also like the parameters to have clear
physical meanings, such as masses and mixing angles of physical
particles.

\subsection{Results}

These experiments were interpreted in a global analysis by Amaldi
{\it et al.} \cite{amaldi} in 1987, in a theoretical collaboration which
consisted of three theorists and five experimenters.  There was
no evidence for any deviation from the standard model: it
is correct to first approximation, and many
contrived models with unusual values for the gauge boson masses
were eliminated.  The model independent analyses were repeated
and improved.  There were unique values for the 4-Fermi
parameters relevant to $\nu q$, $\nu e$, and $eq$ scattering, as
can be seen in Figures~\ref{fig5}, \ref{fig6}, and \ref{fig7}.
These also featured considerably smaller error bars than in the
previous analyses (Figures~1--3 and Table~1.)

\begin{figure}
\postscript{xxnq.ps}{0.7}
\caption[]{Current allowed regions for the neutrino quark couplings,
including results published since \protect\cite{amaldi}.}
\label{fig5}
\end{figure}
\begin{figure}
\postscript{xxnue.ps}{0.8}
\caption[]{Current status of the $\nu e$ parameters, including the
recent CHARM-II results~\protect\cite{CII}.  The $\nu_\mu e$ data
allow four solutions, which differ by the interchange of $V$ and
$A$ and an overall sign difference.  The $\nu_ee$ data eliminates
two of these solutions.  The vector-dominant solution is
eliminated by $e^+ e^-$ annihilation data under the additional
(by now plausible) assumption that the interaction is dominated
by the exchange of a single $Z$ boson.}
\label{fig6}
\end{figure}
\begin{figure}
\postscript{xxeq1.ps}{0.8}
\caption[]{Regions of the parity-violating $eq$
interaction currently allowed by atomic parity violation, the SLAC
asymmetry, and the combined fit, compared with the predictions of
the standard model.}
\label{fig7}
\end{figure}

The interferences between weak an electromagnetic couplings
observed in $e^+e^-$ and $eq$ processes show that they are not
purely $S$, $P$, and $T$.  (This was already shown in the $eq$
system by the SLAC asymmetry.)  Similarly, the $\nu_ee$
interaction is not $S$, $P$, $T$ because of the observed
interference between charged and neutral currents.  (This also
shows that the $\nu_e e$ interaction is flavor conserving at the
neutrino vertex.  It is the only evidence in that sector \cite{LANL}.)
Strictly speaking, there is no proof that the $\nu q$
interactions are vector and axial, but by applying Occam's Razor
it is reasonable to assume that they are not dominated by $S$,
$P$, and $T$ (especially following the discovery of the $W$ and
$Z$).

By this period it was clear that radiative corrections are
necessary to account for the data, particularly the values of the
$W$ and $Z$ masses compared to the neutral current processes.
The value of the weak angle $\sinn = 0.230 \pm 0.007$ was now
established more precisely than before, with the uncertainties
now mainly theoretical, dominated by uncertainties in the $c$
quark threshold in deep inelastic scattering.\footnote{The
uncertainty is mainly in charged current scattering, but the
relevant quantity is the ratio of neutral to charged current
cross sections.} From the collection of data one could set a
robust upper limit $m_t < 200 \gev$ on $m_t$.

Another consequence of this generation of experiments, including
neutral current, charged current, and gauge boson properties, was
that (assuming the standard model gauge group and reasonable
assumptions) the fermion
assignments of all of the new fermions could be determined
uniquely~\cite{unique}.  That is, the left-handed particles occur
in doublets and the right-handed particles in singlets.  In
particular, measurements of the weak interactions of the $b$
imply that the $t$ exists.  This is a compelling argument based
directly on experimental data, which complements theoretical
arguments involving anomalies.  Similarly, the properties of the
$\tau$ imply that the $\nu_\tau$ must exist.

The more precise coupling constants allowed a cleaner test of
grand unification, as can be seen in Figures~\ref{fig4} and
\ref{fig8}.
\begin{figure} 
\vspace{11cm}
\caption[]{Predictions of ordinary and supersymmetric grand unified
theories for $\sinn$ and $m_t$ compared with the data in 1987.
Based on the
analysis in \protect\cite{amaldi}.}
\label{fig8}
\end{figure}
The results again show that ordinary $SU_5$ and similar models
are excluded, while the supersymmetric extensions are in good
agreement with the data, ``consistent with SUSY GUTS and perhaps
even the first harbinger of supersymmetry''\cite{amaldi}.

Finally, there were stringent limits placed on many types of new
physics during this period, such as the masses and mixings of heavy
$Z'$ bosons, the mixings of exotic fermions with unusual weak
interactions, exotic Higgs fields (Figure \ref{fig9}),
leptoquarks,  and 4-Fermi
operators.
\begin{figure} 
\vspace{10cm}
\caption[]{Constraints on $\rho_0$ vs $\sinn$ in 1987, from
\protect\cite{amaldi}. $\rho_0$ is predicted to be one in the
minimal standard model but could differ from unity in models with
Higgs triplets, \etc.  The different constraints from the various
experiments illustrate the power of a global analysis in
giving more stringent results than any one experiment. The curves
shown assume $m_t < 100$ GeV. For more recent results and the
effects of a larger $m_t$, see \cite{PLML}.}
\label{fig9}
\end{figure}

\section{The LEP Era}

LEP has been running since 1989, bringing us into a much more
precise era of precision tests.  Typical results are in the 0.1\%
range, where the radiative corrections are essential.  One
stringently tests the standard model, constrains $m_t$,
and searches for small perturbations due to new physics.

\subsection{The Experiments}

The four LEP experiments -- ALEPH, DELPHI, L3, and OPAL -- have
measured the $Z$ mass to the amazing precision of $M_Z = 91.187
\pm 0.007 \gev$, and have made excellent measurements of the
various widths and asymmetries, such as $\Gamma_Z$, $\Gamma_{f
\bar{f}}$, $A_{FB}(f)$, $A_{pol}(\tau)$~\cite{LEP}.  Recently the
SLD collaborations~\cite{SLD} at the SLC have made the first
measurement of the polarization asymmetry $A_{LR}$.  Many of the
current observables are shown in Table~2          along with
their standard model predictions.  There are also new precise
measurements of $M_W$ from CDF~\cite{CDF}, $M_W/M_Z$ from
UA2~\cite{UA2}, the weak charge $Q_W$ in cesium~\cite{cesium},
and a new generation of $\nu_\mu e$ scattering from CHARM~II~\cite{CII}.

 \begin{table}
\centering
\small
\begin{tabular}{|l|l|l|} \hline
Quantity & Value & standard model \\ \hline
$M_Z$ (GeV) & $91.187 \pm 0.007$ & input   \\
$\Gamma _Z$ (GeV) & $2.491 \pm 0.007$ & $2.490 \pm 0.001 \pm 0.005
                                      \pm [0.006]$\\
$ R = \Gamma_{had}/ \Gamma_{l \bar{l}}$ & $20.87 \pm 0.07$ & $20.78
\pm  0.01 \pm 0.01 \pm [0.07]$ \\
$\sigma^h_p (nb)$ & $41.33 \pm 0.18$ & $41.42 \pm 0.01 \pm 0.01 \pm
                                                [0.06]$\\
$\Gamma_{b \bar{b}}$ (MeV) & $373 \pm 9$ & $375.9 \pm 0.2 \pm 0.5 \pm [1.3]$ \\
$A_{FB} (\mu)$ & $0.0152\pm 0.0027$ & $0.0141 \pm 0.0005 \pm 0.0010$
\\
$A_{pol} (\tau)$ & $0.140\pm 0.018$ & $0.137 \pm 0.002 \pm 0.005$ \\
$A_{e} (P_\tau)$ & $0.134\pm 0.030$ & $0.137 \pm 0.002 \pm 0.005$ \\
$A_{FB} (b)$ & $0.093\pm 0.012$ & $0.096 \pm 0.002 \pm 0.003$ \\
$A_{FB} (c)$ & $0.072\pm 0.027$ & $0.068 \pm 0.001 \pm 0.003$ \\
$A_{LR} $ & $0.100\pm 0.044$ & $0.137 \pm 0.002 \pm 0.005$ \\
                                                              \hline
$\Gamma_{l \bar{l}}$ (MeV) & $83.43 \pm 0.29$ & $83.66 \pm 0.02 \pm 0.13$ \\
$\Gamma_{had}$ (MeV) & $1741.2 \pm 6.6$ & $1739 \pm 1 \pm 4 \pm [6]$ \\
$\Gamma_{inv}$ (MeV) & $499.5 \pm 5.6$ &
$500.4 \pm 0.1 \pm  0.9$ \\
$N_{\nu}$ & $3.004 \pm 0.035 $ & $3$ \\

$\bar{g}_A$ & $-0.4999 \pm 0.0009$ & $-0.5$\\
$\bar{g}_V$ & $-0.0351 \pm 0.0025$ & $-0.0344\pm 0.0006 \pm 0.0013$
\\
$\bar{s}^2_W \ (A_{FB}(q))$& $0.2329\pm 0.0031$ & $0.2328 \pm 0.0003
\pm 0.0007 \pm$ ? \\
  \hline
$M_W$ (GeV) & $79.91 \pm 0.39$ & $80.18 \pm 0.02 \pm 0.13$ \\
$M_W/M_Z$ & $0.8813 \pm 0.0041$ & $0.8793 \pm 0.0002 \pm 0.0014$ \\
$Q_W (Cs)$ & $-71.04 \pm 1.58 \pm [0.88]$ & $-73.20 \pm 0.07
                                                    \pm 0.02$ \\
$g_A^e (\nu e \RA \nu e)$ & $-0.503 \pm 0.017$ & $-0.505 \pm 0 \pm 0.001$ \\
$g_V^e (\nu e \RA \nu e)$ & $-0.025 \pm 0.020$ & $-0.036 \pm 0.001
 \pm 0.001$ \\
$\sinn$ & $ 0.2242 \pm 0.0042 \pm [0.0047]$ & $0.2269 \pm 0.0003
                                            \pm 0.0025$ \\ \hline
\end{tabular}
\label{tab2}
\caption[]{Current values of LEP and other recent observables.  Not all
of the LEP observables are independent.  The last column are the
standard model predictions using $M_Z$ as input and assuming the value
and uncertainty in $m_t$ given by the global best fit for $60 \gev <
M_H < 1 \tev$.}
\end{table}

\subsection{Theoretical Inputs}

An enormous effort was needed to accurately calculate the
radiative corrections to $e^+e^-$ annihilation in the vicinity of
the $Z$-pole.  This was carried out by a number of groups, mainly
in Europe.  Much effort has also gone into parametrizations of
small deviations from the standard model.

\subsection{Results}

\begin{figure}
\postscript{xxxq.ps}{0.7}
\caption[]{Values of the weak angle as a function of the approximate
relevant momentum scale for a variety of probes.}
\label{fig10}
\end{figure}

\begin{itemize}

\item Precision standard model tests and $m_t$: there is no
evidence for deviation from the standard model for wide range of
probes and distance scales, indicating that the standard model is
correct down to a distance scale of $10^{-16}$cm (except possibly
for the Higgs sector), as is indicated in Figure~\ref{fig10}.
The radiative corrections are essential for the agreement of the
various observables, indicating that the basic structure of
renormalizable field theory is correct.  The weak angle in the
$\overline{\rm MS}$ scheme is now determined very precisely,
$\sin^2 \hat{\theta}_W (M_Z) = 0.2328 \pm 0.0007$, where the
uncertainty is almost all due to $m_t$.  In the on-shell scheme
the uncertainty, again dominated by $m_t$, is larger, $\sinn
\equiv 1 - M_W^2 / M_Z^2 = 0.2267 \pm 0.0024$.  There is a
fairly precise prediction $m_t = 150^{+19+15}_{-24-20} \gev$
assuming the standard model.  In the minimal supersymmetric
extension of the standard model (MSSM) one has the slightly lower
value $134^{+23}_{-28} \pm 5 \gev$.  The difference is because in
the MSSM there is a light scalar which acts like a standard model
Higgs but which has a relatively low mass.  For most of parameter
space the other supersymmetric particles have no significant role
on the radiative corrections.  The top quark and Higgs mass
constraints are strongly correlated, and to a good approximation
the predicted value is $m_t = \left(150^{+19}_{-24} + 12.5
\ln \left( \frac{M_H}{300} \right) \right) \gev$.  The origin of
the constraints can be seen in Figure~\ref{fig11}.
\begin{figure}
\postscript{xxmt.ps}{0.8}
\caption[]{The extracted values of $\sin^2\hat{\theta}_W$($M_Z$)
from various observables as a function of $m_t$.  They are all
consistent for a top quark mass of around 150~GeV.}
\label{fig11}
\end{figure}
\begin{figure}
\postscript{xxchis.ps}{0.7}
\caption[]{$\chi^2$ distributions for various values of the Higgs mass.}
\label{fig12}
\end{figure}
The $\chi^2$ distributions of the fit to all data are shown for
various values of the Higgs mass in Figure~\ref{fig12}.  Although
the value predicted for $m_t$ is strongly correlated with the
Higgs mass $(M_H)$, there will be no independent significant
constraint on $M_H$ until after $m_t$ is known directly.  The
total $\chi^2$ of the fit varies by only 0.6 as the Higgs
mass ranges from 60 --- 1000 GeV.  However, once $m_t$ is known
independently with reasonable precision, there may be a marginal
constraint on the Higgs mass, at least if $m_t$ is on the low
end, as can be seen in Figure~\ref{fig13}.
\begin{figure}
\postscript{xxhiggs.ps}{0.8}
\caption[]{68\% and 90\% confidence levels on $M_H$ as a function
of the top quark mass, assuming that it has been measured
directly with a precision of 10 GeV.}
\label{fig13}
\end{figure}
\begin{figure}
\vspace{13cm}
\caption[]{Running coupling constants in ordinary and
supersymmetric grand unified theories, from
\protect\cite{polonsky}.}
\label{fig14a}
\end{figure}
\begin{figure}
\vspace{13cm}
\caption[]{Predictions of ordinary and supersymmetric grand
unified theories for $\alpha_s$, compared with various
experimental determinations, from \protect\cite{polonsky}.  The
bands are the experimental average $0.12\pm 0.01$.}
\label{fig14}
\end{figure}

\item As can be seen in Figure~\ref{fig4}, the more precise
coupling constants, especially $\alpha_s$, allow a much more
stringent probe of grand unification.  It is seen in
Figure~\ref{fig14a} that the low energy coupling constants do not
meet when extrapolated assuming the standard model, but they do
meet when extrapolated according to the supersymmetric
extension, suggesting the possibility of some form of
supersymmetric grand unification at a mass scale of some $10^{16}
\gev$~\cite{newgut}.  Instead of just plotting the couplings, one
can use $\alpha + \sin^2 \hat{\theta}_W (M_W)$ to predict the
strong coupling $\alpha_s$ (Figure \ref{fig14}).  One predicts
$\alpha_s = 0.125 \pm 0.002 \; ({\rm input}) \; \pm 0.01 \; ({\rm
theory})$.  The first uncertainty, from the uncertainties in the
input data, is negligible.  The second, which is much larger, is
due to theoretical uncertainties from threshold corrections at
both the low and grand unified scales, and from possible
nonrenormalizable operators~\cite{polonsky}.  This is in
excellent agreement with the present experimental value of
$\alpha_s = 0.12 \pm 0.01 \; (\rm data)$.
However, given the theoretical uncertainties in the
prediction, for this application more precise values of
$\alpha_s$ would not be useful.  One can also apply the more
traditional procedure of using $\alpha + \alpha_s$ to predict
$\sin^2 \hat{\theta}_W (M_W) = 0.2334 \pm 0.0025 \; ({\rm input})
\;\pm 0.003 \; ({\rm theory})$.  This is in excellent agreement
with the experimental value $0.2326 \pm 0.0006 \; (\rm data)$
(this assumes the supersymmetric range for $m_t$).  However,
given the large uncertainty in the input value of $\alpha_s$,
this procedure is less significant than predicting
$\alpha_s$.

\begin{figure}
\vspace{13cm}
\caption[]{Predictions of the ordinary and supersymmetric grand unified
theories for $\sinn$, compared with the experimental value, from
\protect\cite{polonsky}.}
\label{fig16}
\end{figure}

\item One can search for small deviations due to new physics.
The precision tests are sensitive to many types of new physics
into the TeV range.  The model independent analyses of low energy
processes have already been discussed.  There are now more
stringent limits on $\rho_0$ \cite{PLML}.
This could differ from one due to
exotic Higgs representations or in models
with compositeness, while most superstring theories
predict $\rho_0 = 1$.  There are also improved limits on the
mixings of exotic fermions, which are predicted in most $E_6$
models, with ordinary fermions~\cite{london,nardi}.  The LEP data also
allowed much more stringent limits on the mixing of heavy $Z'$
bosons with the ordinary $Z$, though not much improvement on the
masses in the absence of mixing~\cite{newzp}.  There are
stringent limits on new 4-Fermi operators associated, for
example, with compositeness or leptoquarks, especially from
atomic parity violation~\cite{fourf}, and bounds  on the $S$,
$T$, and $U$ parameters~\cite{STU}, which is a parametrization of
types of new physics which only affect the gauge boson
self-energies.  Currently,
\beqa T &=& +0.05 \pm 0.43 \nonumber \\
S &=& -0.29 \pm 0.46 \nonumber \\
U &=& + 0.37 \pm 0.93.\eeqa
$T$ is associated with $SU_{2V}$ (vector) breaking, and manifests itself
in the strengths $G_F^{NC}/G_F^{CC}$ of the neutral current and
charged current amplitudes, and in the $M_W/M_Z$ ratio.  $S$ and $U$
are associated with $SU_{2A}$ (axial) breaking, and are manifested by the
relation between the low energy couplings and the physical gauge boson
masses, $G_F \leftrightarrow M_{W,Z}$.  The current constraints
on $S$ and $T$ are shown in Figure~\ref{fig17}.

\begin{figure}
\postscript{xxst.ps}{.8}
\caption[]{Current regions allowed in $S$ and $T$ from various reactions
at 90\% c.l. The standard model predictions are shown as a
function of $m_t$.}
\label{fig17}
\end{figure}

\item There are various ways to parameterize new physics.  The
$S$, $T$, $U$ formalism is limited to physics which only affects
gauge boson self-energies.  An alternate formalism describes all
types of new physics~\cite{altarelli}, but utilizes only a few of
the observables.  A general formalism is that of deviation
vectors~\cite{LLM}, which is a way of describing all possible
types of new physics and their effects on all observables.  One
defines the component
\beq D_a=\frac{O^{\rm exp}_a - O^{\rm SM}_a (M_Z)}{\Delta O_a} \eeq
of the deviation vector, where $O_a^{\rm exp}$ is the
experimental value of the $a^{\rm th}$ observable and $O_a^{\rm
SM} (M_Z)$ is the prediction of the standard model, using $M_Z$
as input.  The denominator is the total uncertainty, due to the
uncertainties in the experiment, $\Delta M_Z$, the running of
$\alpha$ from $Q^2 \sim 0$ up to the $Z$-pole, $m_t$, and QCD.
If the standard model is correct, the components of the deviation
vector should be in the approximate range $-1$ to 1, while if
there is a deviation from the standard model the direction
(magnitude) of the deviation vector should indicate the type
(strength) of the new physics.  The current situation of the most
precise observable is shown in Figure~\ref{fig18}.
\begin{figure}
\vspace{10cm}
\caption[]{Deviation vectors from many of the most precise
observables (in early 1992).}
\label{fig18}
\end{figure}
In fact, the comparison is too good, suggesting that some of the
experiments may have overestimated their systematic
uncertainties.

\end{itemize}

\section{The Future}

There are many possible future precision experiments.

\subsection{Motivations}

LEP is the most precise facility for electroweak observables.
However, it is sensitive only to the properties of the $Z$ boson,
and is blind to many types of new physics which do not directly
affect the $Z$.  There is a need for other (less precise)
observables which are sensitive to such deviations from the
standard model as $Z'$ bosons or exotic fermions which do not
directly mix with the ordinary particles, new 4-Fermi operators,
or leptoquarks.  Such experiments would be complementary to
direct searches for new particles at high energy colliders and
would be useful for excluding possibilities even if no deviations
are observed.

\subsection{Possible Experiments}

The LEP program is still in progress.  One can expect improved
values of $\Gamma_Z$, $\Gamma_{f\bar{f}}$, $A_{FB} (f)$, and
$A_{\rm pol} (\tau)$.  Even at present the precision of some of
these observables is better than had every been anticipated
before LEP was built.  It is a remarkable accomplishment, and one
should complete the program.  There may also be precise
measurements of the polarization asymmetry, $A_{LR}$, at SLC and
LEP. $A_{LR}$ is clean theoretically, most systematic
uncertainties cancel, and it is very sensitive to new physics.
However, it is perhaps less crucial than had been anticipated
because of the success of the other LEP observations.

In addition, we anticipate more precise measurements of $M_W$ to
$\sim$ 100~MeV at CDF and D0 and at LEP200.  We also expect major
improvements in the precision of atomic parity violation
experiments at Boulder and Paris.  In the near future the
experiments should improve to the 1\% level, with the uncertainty
dominated by the theoretical matrix elements.  Later, by
comparing various isotopes (for which the atomic uncertainties
largely cancel) one could obtain a precision of $\ll 1\%$.
Atomic parity is very sensitive to some types of deviations,
especially new operators associated with compositeness.  There is
also a possibility of a new deep inelastic $\nu N \ra \nu X$
experiment at Fermilab.  This would be at higher energies than
before, and many of the theoretical uncertainties would be
smaller.  Finally, there may be precision experiments at HERA,
there are proposals for new generations of $\nu e$ and $\nu
p$ scattering at LANL, and there may be $e^{\uparrow \downarrow}
N$ experiments at CEBAF, BATES, and possibly SLAC.  It is not
clear whether the latter will be used more as a test of the
standard model, or (assuming the standard model) as a probe of
nucleon.

\subsection{Implications}

Some time ago Luo, Mann, and I~\cite{LLM} became interested in
the relative sensitivity of these different proposals, {\it
i.e.}, to what extent would each be sensitive to various types of
new physics, which would be more sensitive, and to what extent
would they be complementary.  We carried out an analysis of many
proposed types of new physics and many observables, calculating
the expected deviation vectors for some 30 types of new physics
for a number of possible observables.  A typical example is shown
in Figure~\ref{fig19}.
\begin{figure}
\vspace{13cm}
\caption[]{Sensitivities of a number of types of observables to a
particular type of new heavy $Z$ boson.  The larger bars
represent a higher sensitivity, shown in GeV in the right-hand
column.  For this particular example, $M_W$, deep inelastic
neutrino scattering, atomic parity violation, and various LEP
asymmetries are all very sensitive.}
\label{fig19}
\end{figure}

The conclusion was that there is no one ``best'' observable.  A
number of proposed experiments are important, depending on the
type of new physics.  A new program of precisions experiments
would give an unprecedented test of the structure of
renormalizable field theory and would be sensitive to small
deviations from the standard model.  It is important that as many
as possible of these experiments be done to maximize the
sensitivity to different types of new physics; to simultaneously
determine the parameters of the standard model; to confirm any
discrepancies; and to diagnose the origin of discrepancies ({\it
i.e.}, the patterns of deviations for the various experiments are
quite different for different types of new physics).  To carry
out this program it will be important to have a uniform analysis
of all of the experiments with consistent definitions of $\sinn$.
At present the situation is confused because different
definitions are being used, some of which are effective
parameters defined by complicated computer programs.  This is
adequate as long as one is just treating the LEP data.  However,
as soon as one tries to compare to other observables or to such
theoretical ideas as grand unification the uncertainties in the
meaning of these effective parameters become a problem.  It
would be desirable to use standardized definitions of parameters,
such as the $\overline{\rm MS}$ definition of the weak angle.  It
is also crucial that careful estimates of systematic
uncertainties and correlations be applied to the data.

\section{Conclusions}

\begin{itemize}

\item The weak neutral current plus the $W$ and $Z$ properties are the
primary test of electroweak unification.

\item These observables have uniquely established the standard model
gauge group and fermion representations.

\item The standard model is correct to an excellent first
approximation down to a distance scale of $10^{-16}$~cm.

\item Amongst the results
\begin{itemize}

\item In the $\overline{\rm MS}$ scheme $\sin^2\hat{\theta}_W (M_W) =
0.2328 \pm 0.0007$.

\item In the on shell scheme $\sinn \equiv 1 - M_W^2/ M_Z^2 =
0.2267 \pm 0.0024$.  The uncertainties in both cases are largely
due to $m_t$.

\item One has the predictions $m_t = 150^{+19+15}_{-24-20} \gev$ in
the standard model and $134^{+23}_{-28} \pm 5 \gev$ in the MSSM.

\item  Although the values of $M_H$ and $m_t$ are strongly correlated
there is no significant $M_H$ constraint until $m_t$ is known
directly.

\item  The observed coupling constants are in remarkable agreement
with the predictions of supersymmetric grand unification.  This could
well be an accident, but on the other hand it may be pointing us
toward supersymmetric unification and possibly superstring theories.
\end{itemize}
\item The structure of gauge field theory is confirmed.

\item The observables are sensitive to many types of new physics into
the TeV range  and are an excellent complement to high energy
colliders.

\item Let us not cut short the program of LEP and other precise
measurements; it is unique in the history of particle physics and
should not compromised.

\item It is important to present all results in terms of
well-defined quantities, such as a consistent definition of
$\sin^2\hat{\theta}_W$, to compare the results of different
classes of experiments and with the predictions of grand unified
theories.

\end{itemize}

\end{document}